\documentstyle{article}        
\newcommand{\bib}{\bibitem}
\newcommand{\eqr}[1]{ (\ref{#1})}

\newcommand{\spz}{\hspace{0.7cm}}

\newcommand{\nn}{\nonumber} 

\newcommand{\de}{\partial} 
\newcommand{\ri}{\right} 
\newcommand{\lf}{\left}

\newcommand{\eq}{\begin{equation}}
\newcommand{\en}{\end{equation}}
\newcommand{\bea}{\begin{eqnarray}}
\newcommand{\eea}{\end{eqnarray}}

\newcommand{\ba}{\begin{array}}
\newcommand{\ea}{\end{array}}

\newcommand{\virg}{\spz,}
\newcommand{\pu}{\spz.}

\newcommand{\half}{\frac{1}{2}}

\newcommand{\M}{{\hbox{\large $\cal M \;$}}}

\newcommand{\T}{{\hbox{$Tr\;$}}}

\newcommand{\SLd}{$SL \lf ( 2, {\bf R} \ri )$ }

\newcommand{\SUU}{$SU \lf ( 2 \ri )/ U\lf ( 1\ri )$ }
 
\newcommand{\sm}{ $\sigma $- model }
\newcommand{\sms}{ $\sigma $- models }
\newcommand{\AP}[1]{{\it Ann.\ Phys.}\ {\bf #1}}
\newcommand{\CMP}[1]{{\it Comm.\ Math.\ Phys.}\ {\bf#1}}

\newcommand{\NP}[1]{{\it Nucl.\ Phys.}\ {\bf #1}}
\newcommand{\PL}[1]{{\it Phys.\ Lett.}\ {\bf #1}}

\newcommand{\PRL}[1]{{\it Phys.\ Rev.\ Lett.}\ {\bf #1}}
\newcommand{\TMP}[1]{{\it Theor.\ Math.\ Phys.}\ {\bf #1}}
\begin{document}

\setlength{\unitlength}{.8mm}

\hfill  DFUL-1/02/97\\ \vskip 1cm
   \begin{center}  {\large \bf Integrable dissipative structures
 in the  gauge theory of gravity } \\[.5cm]
\vskip 1.2cm 
 {\large  \it L.  Martina \footnote{E-mail: martina@le.infn.it}
             , O.K. Pashaev$^{\dagger \,*}$ \footnote{E-mail:
pashaev@main1.jinr.dubna.su} and G. Soliani \footnote{E-mail:
soliani@le.infn.it}}\\
     
\vskip .8cm
  {\it Dipartimento di Fisica dell'Universit\`a and INFN - Sezione di 
 Lecce}\\
 {\it 73100 Lecce, Italy}\\ $\dagger)$ {\it Joint Institute for Nuclear
Research, 141980 Dubna, 
 Russia}  \footnote{ Permanent address} \vskip 1cm 
\noindent {\bf Abstract} \\
\end{center}
\noindent { The  Jackiw-Teitelboim  gauge formulation of the
 1+1 dimensional gravity allows us  to  relate different gauge fixing
conditions with integrable hierarchies of evolution equations.  We show that
the  equations for the   Zweibein fields can be written as a pair of  time
reversed evolution equations of the reaction-diffusion type, admitting
dissipative solutions. The spectral parameter for the related Lax pair
 appears as the constant valued spin connection associated with the $SO(1,1)$
gauge symmetry. Spontaneous breaking of the non-compact symmetry and the
irreversible evolution are  discussed.\par \bigskip \noindent
PACS numbers: 04.70.Dy, 11.15-q, 11.10.Lm }
\newpage
\section{}  In  the last few years it has been shown that the classical 
treatment of the topological field theories can be strictly related to 
completely integrable models, which  realize special
non-covariant gauge conditions. At the classical level the topological field
theories have zero Lagrangians on the solution manifold and the
corresponding  equations  of motion coincide with the zero field strenghts.  In a
geometrical language this fact corresponds to the zero curvature condition
required for the integrability of certain affine connections (the so-called
linear problems),  associated with the completely integrable systems. 

On  the other hand,  it has been developed  a  systematic procedure in order 
to associate a  completely integrable model with a Chern-Simons (CS)  gauge
theory \cite{LP,MPS1,MPS2}. This  procedure has  been  performed for a
compact group, but  most of the  involved  algebra can be   reused in the 
non-compact case. However, one can expect that the resulting nonlinear
models  may    display some interesting    physical  and  mathematical
features. Indeed,  from one side the non-compact groups appear tipically
in  gravity theories   as  global symmetry  groups  and local
symmetries  in  their gauge  formulations.  Furthermore, one  knows that
unstable and even chaotic  dynamics is related  to the hyperbolic  and
non-compact geometrical structures \cite{Arn}.  In the present   work we consider 
the model called   ``  lineal  gravity'', introduced    by   Jackiw
\cite{J} and Teitelboim \cite{T}, in the form of a  \SLd gauge theory in $\lf(
1 +  1  \ri)$ dimensions of the BF type \cite{IT,CW,MS,BBRT}.  This  model  is equivalent  to  the Liouville gravity,     whose  
quantization  has     been  discussed   recently
\cite{Seib}. In   spite of its    simple  physical content, we are
considering  this theory as   a  mathematical laboratory  in  order to
   describe  explicitly the abovementioned embedding.  Although the
conformal symmetry is  generically  broken,  the  integrable  character  of 
some  of the resulting equations  may provide  a  more  direct application of
the  quantum inverse   scattering. Furthermore, we   show   that 
the equations for  the Zweibein fields  are given  by  the pair of time
- reversed nonlinear equations as in  the  termo-field theory  \cite{Cel}.
  In fact, for the  particular gauge  condition 
provided by the 
$O(2,1)$  Heisenberg   model,  a system   of reaction- diffusion equations
arises,  which appears in the theory  of chemical reactions and in dissipative
structures.\par From the point of view of a gauge field theory of 
gravitation, the main idea is that the general   coordinate transformations  
are  implemented by   gauge ones. Thus, the gauge  groups induce a metric 
on a base space. On the other  hand,  one can  always build   up  a
non-compact   nonlinear \sm associated with  certain zero-curvature 
conditions, resorting  to the so-called moving trihedrals    \cite{Car, Orfa}. In
the   following we shall study the class  of \sms related to  the \SLd group.
The algebra of  this group and the corresponding  Killing metric is given in
terms of the basis elements $\tau_{i}$ ( $ i = 0, 1, 2$) by the bilinear
product \eq \tau_{i} \tau_{j} = h_{i j} + i\, c_{i  j k} \tau_{k}, \en where $\lf ( 
h_{i j}\ri)={\rm  diag}(-1,-1,1)$  and $c_{i  j k} =   -
\epsilon_{i j l} h_{l k}$. With any point of  a given base space \M endowed
with the non positively definite  \SLd metric we associate an orthonormal
frame by  a local
 adjoint representation of the algebra,  expressed by \eq  g \tau_{i} g^{-1} = {
n}^{k}_{i}\tau_{k}.
\en The  triad of vector fields  ${\bf  n}_{i}\lf(x\ri)= \lf( h_{j j}{ n}^{i}_{j}\ri)$
would satisfy  the orthonormal conditions \eq \lf({\bf n}_{i}{\bf n}_{j}\ri) =
h_{ij} ,\,\,\, {\bf n}_{i}\wedge {\bf n}_{j} = c_{ijk} {\bf n}_{k}.\en  If $g$ is a 
differentiable  function on \M, the local frame $\lf\{{\bf n}_{i}\ri\}$ changes
accordingly to the adjoint representation  of  the right-invariant   chiral
currents  $J_{\mu}  = g^{-1}\de_{\mu}g$, that is   \eq \partial _{\mu }{\bf 
n}_{i}= (J_{\mu })^{(ad)}_{ik}{\bf n}_{k} , 
\label{ntra}\en where $J_{\mu}^{(ad)}$  are matrices in the adjoint
representation.  Equations \eqr{ntra} can   be  seen as a   linear system   for
the triad $\lf\{{\bf  n}_{i}\ri\}$.  Its integrability is assured by the zero
curvature  condition, satisfied by the  chiral currents $J_{\mu}  $ over a  
simply connected base space. Now we assign  to one  of the  vector fields,   
say
${\bf n}_{0}=  {\bf s}$, the special role of (pseudo)-spin variable. It has  the 1-sheeted hyperboloid   $\lf( {\bf  s}, {\bf s}\ri) 
= -1$ as
phase space. The vector fields ${\bf  n}_{1}$ , ${\bf n}_{2}$  describe  the
tangent   plane   of such  a  hyperboloid  and      can  be locally rotated.
In  other words a residual local  $SO \lf( 1, 1\ri)$ symmetry survives,
corresponding to  local Lorentz transformations. In order to specify
this aspect, let  us introduce a new basis  ${\bf n}_{\pm}   =   {\bf  n}_{1}
\pm   {\bf   n}_{2}$   and  the  following parametrization of the chiral
currents \bea J_{\mu} = {i \over 4} \tau _{0}V_{\mu }  + \lf({\matrix{ 0& 
q_{\mu}^{-}\cr  q_{\mu}^{+}& 0 \cr}}
\ri),\label{curr}\eea  where  $V_{\mu}$ and $q_{\mu}^{\pm}$   are real 
quantities defined by \eq V_{\mu } = 2({\bf n}_{2}, \partial _{\mu }{\bf
n}_{1}), \qquad q^{\pm}_{\mu }  =   \pm  {1  \over 2}  ({\bf    s},
\partial_{\mu}{\bf n}_{\pm}).  \en  This   parametrization  realizes a   ${\bf
Z}_{2}$  - graduation of  the gauge algebra with  isotropy group ${\bf R}$.
Then,  the system \eqr{ntra} takes the form \bea & D^{\mp}_{\mu }{\bf
n}_{\pm}& =
\mp  2 q^{\pm}_{\mu } {\bf   s} ,\nn \\ &   \partial _{\mu }{\bf s} &=
q^{+}_{\mu } {\bf n}_{-} - q^{-}_{\mu } {\bf n}_{+} ,
\label{linsys}\eea where $D^{\pm}_{\mu } \equiv  \partial _{\mu } \pm (1/2)
 V_{\mu }$ is the covariant derivative. This   form   is  
 invariant under the local $SO(1,1)$ gauge transformations generated by an
arbitrary real function $\alpha$ \eq {\bf s} \rightarrow {\bf s}
 ,\; {\bf n}_{+}\rightarrow e^{+\alpha }\, {\bf n}_{+} , {\bf n}_{-}
   \rightarrow e^{-\alpha } \,{\bf n}_{-} , \en or equivalently \eq
 V_{\mu }'= V_{\mu } + 2 \partial _{\mu }\alpha  ,\, q^{+}_{\mu }\;'= e^{\alpha
} q^{+}_{\mu },\, q^{-}_{\mu} \; '= e^{-\alpha } q^{-}_{\mu },  \en which are just
the local boost rotations in the tangent plane to the hyperboloid described by 
{\bf s}.  In this formalism,  the zero curvature condition becomes \bea 
D^{\mp}_{\mu }q^{\pm}_{\nu } &=& D^{\mp}_{\nu
         }q^{\pm}_{\mu }  , \nn \\ \partial_{\mu}V_{\nu} -
  \partial_{\nu}V_{\mu} &=& 4 (q^{+}_{\mu }q^{-}_{\nu } - q^{+}_{\nu
}q^{-}_{\mu }) \label{eqm} .\eea This system can be written in a base space of
arbitrary dimension. However, it is natural to consider it in
  2 or 3 dimensions. In the latter case Eq. \eqr{eqm} represents the Euler -
Lagrange equations for the CS action on the \SLd group \cite{Witt} , which
 can be seen as a subgroup of the corresponding Poincar\'e group. On the
other hand, in 2 - dimensions the local symmetry group identifies with the de Sitter
group. In this framework one can introduce  the action \bea S
&= &{k \over 2\pi}
\int_{\M} \T \lf( J_{0} \, F_{1  2}\ri) \, dx^{1}\,
    dx^{2}  \label{BF} \\ &=& {k \over 2\pi} \int_{\M} \epsilon_{a \, b} \lf [
q_{0}^{+}{ D}^{+}_{a}{ q}^{-}_{b} + q_{0}^{-}{ D}^{-}_{a}{ q}^{+}_{b}
 + {1 \over 8} V^{0}({\de_{a} V_{b}} - 4 { q}^{+}_{a}{ q}^{-}_{b})\ri]\,
   dx^{1}\, dx^{2},\nn\eea where $ F_{1  2} = {\partial
 }_{1}{J}_{2}-{\partial }_{2}{J}_{1}+\left[{{J}_{1},{J}_{2}}\right]$, $\epsilon_{a
b}$ is antisymmetric ($a, b =1, 2$)  and $J_{0}$ plays the role of Lagrangian
multiplier.  The  field
   equations  are $F_{1  2} = 0$ and ${\cal D}_{a} J_{0} = 0$, where ${\cal
D}_{a} = \de_{a} -\lf[ J_{a}, \;\;\ri ]$.  They are  given  equivalently by 
 \eqr{eqm} for  $\mu, \nu = 0, 1, 2$, under the assumption that all the 
derivatives of $q^{\pm}_{\mu}$  and  $V_{\mu}$  with
 respect to $x^{0}$  vanish.  This structure is a relict
       of the 3 - dimensional  CS theory under the effects of the dimensional
reduction.   Of course, the space of all local
       solutions of the classical field equations, modulo gauge
               transformations, is finite dimensional. Furthermore,
${\bf s} \lf(x^{0}, x^{1}, x^{2}\ri)$ and ${\bf n}_{\pm} \lf(x^{0}, x^{1},
x^{2}\ri)$ can be  found, by  solving  the linear  differential
equations \eqr{linsys} with
$x^{0}$-independent
 coefficients.  In particular, for $\mu = 0$, the
   eigenvalues of this  system are $  0$ and $ \pm 2
   \sqrt{\lf(J_{0}, J_{0}\ri)}$. Remarkably, $\de_{\mu} \lf(J_{0}, J_{0}\ri) = 0$. 
Thus,  instead of the variables ${\bf s}$ and ${\bf n}_{\pm}$ one could use
$J_{0}$ and
$\de_{a} J_{0}$
$\lf (a =1, 2\ri )$ as a local frame,  treating in this framework the system in
analogy with 
\cite{AER}. 
     Action \eqr{BF} allows  us to take contact with the 2 - dimensional  Jackiw
- Teitelboim gravity model \cite{J,T}, by introducing the metric tensor  \eq
g_{a
         b} = 4  (q^{+}_{a}q^{-}_{b} + q^{+}_{b}q^{-}_{a}) =
    e_{a}^{m}e_{b}^{n}\eta_{m n} \virg \label{metric}\en where the
   Zweibein and the spin connection are given by  \eq e^{0}_{a} \pm
 e^{1}_{a} = 2 \sqrt{2} q^{\pm}_{a}, \qquad \omega_{a} = \half  V_{a}
\en and the local flat metric is $\eta_{m n} = diag(-1, 1)$. In these settings 
Eq. \eqr{eqm} prescribes the constancy 
 of the 2 - dimensional scalar curvature 
 to the value of the cosmological constant,   normalized to  $-1$, and
the
   torsionless condition of the world-sheet manifold. Indeed,  Equations 
   \eqr{eqm} can be derived also from the action  \cite{J,T} \eq S=
      \int_{\M}  {\sqrt{ -g\;}} \; V_{0}  (R + 1) dx^{1}\, dx^{2} \virg
  \label{JT}\en where $V_{0}$ is   a world  scalar 
 Lagrange multiplier   related
to the dilaton field. Action \eqr{JT} admits 
as local symmetry group the de Sitter
 group \SLd, with local Lorentz generator $L = {i \over 2} \tau_{0}$.
Furthermore, the gauge invariance of the  theory  based on \eqr{BF} is
equivalent
   to the general covariance  of \eqr{JT} on the solutions of Eqs. 
  \eqr{eqm}. Finally, the metric tensor $g_{a b}$ has the direct and
  simple  representation in terms of spin variables \eq g_{a b} =  2
\lf(\de_{a} {\bf s}, \de_{b} 
{\bf s}\ri) \pu \label{Smetric}\en This formula enables us to give
a gravitational interpretation of the \sm  we are going to introduce.
In particular, Equation \eqr{Smetric} expresses that 
the (pseudo)-spin $\bf s$ provides a 
special parametrization of the surface manifold.
\section{} The idea we follow is to add  to \eqr{linsys} a  differential
constraint  in the $\lf( x^{1}, x^{2} \ri)$ space for $\bf s$, such that
 a completely integrable dynamics is introduced in order to partially fix the
gauge freedom in a controlled fashion and allowing a residual local Lorentz
covariance.  Moreover, we concentrate on the "effective" space-time $\lf
(x^{1}, x^{2} \ri)$, forgetting for the moment all what concerns the 
    variable $x^{0}$ and the current $J_{0}$.  Precisely, we consider as a
 constraint the classical continuous Heisenberg model realized on the
\SLd$/SO\lf( 1, 1\ri)$ coset space  \eq \partial _{2}{\bf s} = {\bf s \wedge
    \partial }^{2}_{1}{\bf s} \label{HM}\pu \en Substitution into
       \eqr{linsys} leads  \eq q^{+}_{2} =
  D^{-}_{1}q^{+}_{1} ,\qquad  q^{-}_{2} = - D^{+}_{1}q^{-}_{1} .\label{constr} \en
  These relations    allow us to exclude the $q^{\pm}_{2}$ from the
 fields  equations \eqr{eqm},  obtaining  \bea D^{\mp}_{2}q^{\pm}_{1}
\mp (D^{\mp}_{1})^{2}q^{\pm}_{1} = 0,\\ \partial _{2}V_{1} - \partial
  _{1}V_{2} = 4\partial _{1}(q^{+}_{1}q^{-}_{1}) .\eea Defining the
   flat connection \eq A_{2} = V_{2} + 4(q^{+}_{1}q^{-}_{1} -
\rho),\qquad  A_{1}= V_{1},\label{irro}\en where $\rho$ is an arbitrary
real constant,  and gauging out it by a local
$SO
\lf(
 1, 1 \ri)$ transformation  $A_{j} = 2\partial _{j} \lambda , \qquad q^{\pm}  =
q^{\pm}_{1} e^{\pm\lambda }$ for any regular real function
  $\lambda$, we get    the  nonlinear reaction-diffusion system \eq
\partial _{2}q^{\pm}  \mp  \partial ^{2}_{1}q^{\pm} \pm  2(q^{+}q^{-}
 - \rho)q^{\pm}  = 0 \pu  \label{readiff}  \en Here only the   global
   $SO \lf( 1, 1 \ri)$  $q^{\pm}\rightarrow e^{\pm \alpha} q^{\pm}$
  invariance survives.  The system \eqr{readiff} represents a particular form
of a 2 - component reaction - diffusion system, playing  an important role in
synergetics  \cite{Pri,Hak,Koch}.
\par The equations  \eqr{readiff} come from the    Lagrangian density
\eq{\cal  L} = {1 \over 2}(q^{+}\partial_{2}q^{-} - q^{-}\partial_{2}q^{+}) -
    \partial_{1}q^{+}\partial_{1}q^{-} - (q^{+}q^{-})^{2}.\en  By resorting to  the
first order formalism \cite{FaJa}  we obtain the canonical pairs
   \eq \{q^{-}(x), q^{+}(y)\} = \delta \lf (x - y\ri) \pu \en  The
     Hamiltonian density is   \eq H = \partial_{1} q^{+}
\partial_{1} q^{-} + (q^{+}q^{-})^{2}\virg\en whose integral provides
 the total conserved energy. 
 The conserved quantity associated with  the
 hyperbolic scale invariance of the system \eqr{readiff} is $M = \int
q^{+}(x)q^{-}(x) dx$. We will see below that  M is interpreted as the
  total mass of a given configuration. It should be noticed that the mass
integral is  related  to the energy of  the Heisenberg model \eqr{HM}
  by ${\cal E} = \int \partial_{1}{\bf s} \partial_{1}{\bf s} = 4M$.
        Furthermore, the corresponding momentum is  $P = \int
        (q^{+}\partial_{1}q^{-} - q^{-}\partial_{1}q^{+})dx$. 
Actually,  we will show that the system
 \eqr{readiff} is completely integrable and it admits infinitely many integrals
in involution.  This property comes from the relation with the
  model \eqr{HM}, for which we can  write down the Lax pair
  representation. This phenomenon is well studied in the case of the
     compact  \SUU continuous Heisenberg model,  where the
    correspondence is established with the nonlinear Schr\"odinger equation
\cite{ZakTa,LP}. 
\par In the gauge  fixed by \eqr{HM}  the metric tensor \eqr{metric} takes
the form 
\bea g_{00}  &=& - 8 \partial_{1}q^{+}\partial_{1}q^{-},\nn \\ g_{11}
     & =& 8 q^{+}q^{-}, \label{metricHM} \\ g_{01} &=& g_{10} = 4
(q^{-}\partial_{1}q^{+} \, - \, q^{+}\partial_{1}q^{-}).\nn \eea From the 
comparison of  these formulas with the conserved quantities shown above, 
we observe that the components $g_{11}$ and $g_{01}$ have  the
    meaning of the mass and the momentum densities, respectively.
\par A simple symmetry analysis of Eq. \eqr{readiff} provides
 the time-inversion symmetry
$q^{+} \rightarrow q^{-}, q^{-} \rightarrow q^{+}, t \rightarrow -t$ and
 the space reflection symmetry $x \rightarrow -x$.  By inspection one finds
also  the Galilei invariance 
  $  x \rightarrow   x + 2 v t, \; t \rightarrow t, \;
 q^{\pm} \rightarrow
       e^{\mp\alpha(x',t')} q^{\pm}(x',t') $, where  $
 \alpha(x,t)  =  v^{2}t + v x$ . 
\par However, the full symmetry structure of the system \eqr{readiff} is
encoded into  the Lax pair representation,  which can be obtained by
following the procedure adopted  in \cite{LP}.  The  main idea is  to reduce
the space  of the   chiral   currents \eqr{curr}  consistently   with  the
constraints
\eqr{constr} and provide a grading,  by exploiting the freedom in  the 
choice of the vector  field  $V_{\mu}$.  In fact,  we first  substitute  the 
relations  \eqr{constr} and   \eqr{irro}  into the expression \eqr{curr}  of
$J_{a}$ ($a   =  1, 2$). The  corresponding  consistency condition   provides a
system   of differential  equations a  little  more general   than  \eqr{readiff},
because  of    the explicit presence of
$A_{a}$. However,  we can gauge  out these fields in terms
of an arbitrary scalar  function.  At least  locally, we can set
$A_{2}  = - \half A_{1}^{2}  = - 8  \zeta^{2}$,  where $\zeta$ is a real
constant.  A    further Galilei transformation  furnishes  the Lax pair 
for  the system  \eqr{readiff}  \bea L_{1} &= & \de_{1}  +
\left({\matrix{ \zeta  &q^{-}\cr q^{+} & -\zeta
\cr}}\right) ,  \nn \\  \label{lax} \\     L_{2} & =&\de_{2} +
   \left({\matrix{   2 \zeta^{2}   - (q^{+} q^{-}      -  \rho) &-(\partial_{1} 
 -  2 \zeta)q^{-}\cr   (\partial_{1} +   2 \zeta)q^{+}  & - 2 \zeta^{2}   + (q^{+} q^{-}  -  \rho)  \cr}}\right). \nn \eea Thus, we  have  associated the
spectral  parameter $\zeta$  of the Lax pair with  a constant  value of the 
$SO(1,1)$  spin connection. The equations $L_{1} \Psi = 0$ and  $L_{2} \Psi = 0$ provide the auxiliary linear problems, whose integrability is assured by  \eqr{readiff}.  In the compact $U(1)$ gauge theory the 
analogous situation is interpeted as the value of   the condensate state   of
the  statistical gauge  field
\cite{LP}.   The Lax  pair  \eqr{lax} is  of  the Zakharov-Shabat type
\cite{ZakSha}, where a rotation of $\pi/2$ in the complex plane of the spectral
parameter  is required.  The B\"acklund
transformations for the system  \eqr{readiff} are
\bea \partial_{1}\lf(q^{\pm} - \tilde q^{\pm}\ri)& =& {\sqrt{\lf(q^{+} - \tilde
q^{+}\ri)\;\lf(q^{-} - \tilde q^{-}\ri) - \mu\;}} \;\lf(q^{\pm} +   \tilde  
q^{\pm}\ri), \label{BTspace}\\  \partial_{2}\lf(q^{\pm} -
\tilde      q^{\pm}\ri)&     =& \pm     {\sqrt{\lf(q^{+}    -   \tilde
q^{+}\ri)\;\lf(q^{-}      -     \tilde     q^{-}\ri)    -        \mu\;}}
\;\partial_{1}\lf(q^{\pm}  +   \tilde q^{\pm}\ri)   \nn  \\  &  \mp\,&
\lf(q^{+}q^{-} + \tilde q^{+}  \tilde q^{-}\ri) \;\lf(q^{\pm} - \tilde q^{\pm}\ri).
\label{BTtime}\eea  Thus,  simple steady  state solutions can be easily  found. 
For  example, for  $\rho =  0$ and the   initial trivial configuration $\tilde q^{+}
= 0 = \tilde q^{-}$ one finds: 
\par\noindent a) for $\mu < 0$,
       \eq   q^{\pm}\lf(x, t\ri) =  {a\, e^{ \pm  \,  a^{2} t} \over \sinh {a\lf(x
-\xi\ri)}} \qquad \lf(\lf|\mu\ri| = a^{2}\ri), \en 
\par\noindent b) for $\mu > 0$,
       \eq q^{\pm}(x,t) = a{e^{\mp \, a^{2}t} \over \cos a(x - \xi)},
 \qquad \lf(\mu = a^{2}\ri)\en with $x = x^{1}$ and $t = x^{2}$. For $a
\rightarrow 0$ one obtains the simple rational  solution
$q^{\pm} = { 1 \over x}$.  In a similar way
one can obtain the steady state periodic solutions, namely
\eq q^{\pm} = b e^{\mp \frac{v}{2} \lf( \xi -
\xi_{0}\ri) } \frac{sn \lf[ a \lf(\xi -\xi_{0}\ri) ; r\ri]} {cn \lf[ a \lf(\xi
-\xi_{0}\ri) ; r\ri]}, \quad {\rm or}\quad q^{\pm} =  \frac{b e^{\mp
\frac{v}{2} \lf( \xi -
\xi_{0}\ri) }} {cn \lf[ a \lf(\xi -\xi_{0}\ri) ; s\ri]}, \en where the moduli $r =
\frac{1}{|a|} \sqrt{a^{2} - b^{2}}$ and 
$s = \frac{|a|}{\sqrt{a^{2} + b^{2}}}$ are given in terms of the real arbitrary
parameters $a, b$. These solutions
present  first order poles on the real axis, located at odd multiples of the
complete elliptic integral of the first kind. Furthermore,  
the system
\eqr{readiff}
$\lf(\rho = 0\ri)$  admits the analogous of  the bright 
 soliton solution   for   the Nonlinear  Schr\"odinger Equation. In a moving
frame coordinate $\xi = x - vt +\xi_{0}$, such a solution is given by  
\eq q^{\pm}(x,t) = \pm k e^{\pm\lf[\lf(k^{2} -  {1 \over 4}v^{2}\ri) t - {1 \over
2}v \lf(\xi - \xi_{0}\ri) \ri]}{\rm  sech}\,  k \xi \pu
\label{brisol}\en It  depends on the two real parameters $k$ and $v$.  But in
contrast to the bright soliton, the
  above solution does not preserve the amplitude  for the  components
$q^{\pm}$ independently. During the time evolution  one of the fields  is
exponentially growing, while the other one is decaying.  At the same time, the
product $q^{+}q^{-}$ has the usual 
 solitonic shape. In the context of synergetics \cite{Pri,Hak,Koch}, where the
dissipative structrures are investigated,  a growing mode is  associated with
 the pattern formation. Consequently, for the solution
\eqr{brisol} the name of "dissipaton"  is suggested.  In some sense it can be
considered as the square root of the soliton. We note that the metric tensor is
a regular bounded function for  the dissipaton solution \cite{last}.
  In the space of
    parameters $\lf(v, k\ri)$ there exists  the critical line   $v =
 2k$. For the  solution \eqr{brisol}  obtained with $v < 2k$, 
  at the  infinity one
has 
$q^{\pm} \rightarrow 0$.  At the critical value $v = 2k$ the solution is  a
steady state  in the moving frame
 $q^{\pm} = \pm k e^{\pm k \xi_{0}} 
\lf(1\, \mp \, \tanh \;k\xi \ri)$, with constant
asymptotics
$q^{\pm} \rightarrow \pm \, 2 k e^{\pm k \xi_{0}}$ for $x
\rightarrow \mp \infty$ and  $q^{\pm} \rightarrow \pm \, 0$ for  $x
\rightarrow \pm \infty$. In the  over-critical  case $v > 2k$,   we are
led to $q^{\pm} \rightarrow \pm \infty$ for $x
\rightarrow \mp \infty$   and  $q^{\pm} \rightarrow \pm \, 0$ for  $x
\rightarrow \pm \infty$ .
 Furthermore, by  computing the first integrals 
                             $M = -2k$, 
                             $P = Mv $ and 
$E = - {1 \over 4}Mv^{2} - {1 \over 12}M^{3} = {1 \over 2}Mv^{2} - {1
  \over 12}M(9v^{2} + M^{2})$,  for $k < 0$ one can  naturally interprete
 $M$ as the mass of a non-relativistic particle with
  momentum $P$ and energy $E$.  The second term in the expression of the
energy  describes
    the bound state energy arising from the nonlinear
  interaction. Actually, by standard techniques (see for instance \cite{Mat}) one finds that  the solution \eqr{brisol} is the first iteration of the $N$-soliton superposition formula
\eq
q^{\pm}\lf[N\ri] = 2 \Delta^{\pm}\lf[ 2N \ri]/ \Delta\lf[ 2N \ri],\en
where 
 \bea  \Delta \left[{2N}\right]=\left|{\matrix{{\psi }_{1}^{+}&.&.&
{\psi }_{2N}^{+}\cr
.&.&.&.\cr
{{{\zeta }_{1}^{N-1}}^{}}^{}{\psi }_{1}^{+}&.&.&
{{{\zeta }_{2N}^{N-1}}^{}}^{}{\psi }_{2N}^{+}\cr
{\psi }_{1}^{-}&.&.&{\psi }_{2N}^{-}\cr
.&.&.&.\cr
{{{\zeta }_{1}^{N-1}}^{}}^{}{\psi }_{1}^{-}&.&.&
{{{\zeta }_{2N}^{N-1}}^{}}^{}{\psi }_{2N}^{-}\cr}}\right|,
{\Delta }^{\pm }\left[{2N}\right]=
\left|{\matrix{{\psi }_{1}^{\pm }&.&.&{\psi }_{2N}^{\pm }\cr
.&.&.&.\cr
{{{\zeta }_{1}^{N}}^{}}^{}{\psi }_{1}^{\pm }&.&.&
{{{\zeta }_{2N}^{N}}^{}}^{}{\psi }_{2N}^{\pm }\cr
{\psi }_{1}^{\mp }&.&.&{\psi }_{2N}^{\mp }\cr
.&.&.&.\cr
{{{\zeta }_{1}^{N-2}}^{}}^{}{\psi }_{1}^{\mp }&.&.&
{{{\zeta }_{2N}^{N-2}}^{}}^{}{\psi }_{2N}^{\mp }\cr}}\right|
\label{determ} \eea and  $\lf(\psi^{+}_{k}, \psi^{-}_{k} \ri)^{T}$
are eigenfunctions  of the auxiliary linear problem generated by  \eqr{lax}
with $q^{\pm} = 0$ and spectral parameter $\zeta_{k}$. The
wavefunction for $q^{\pm}\left[ N\right]$ is given by 
$\Psi\left[ N\right] = \left(1/\Delta\left[2 N\right]\right)
\left(\Delta^{+}\left[2 N + 1\right], 
\Delta^{-}\left[2 N + 1\right]\right)
^{T}$, where
$${\rm \Delta }^{\pm }\left[{2N+1}\right]=
\left|{\matrix{{\Delta }^{\pm }\left[{2N}\right]&{\zeta }^{i}{\psi }^{\pm }\cr
{\zeta }_{k}^{N-1}{\psi }_{k}^{\mp }&{\zeta }^{N-1}{\psi }^{\mp }\cr}}\right|
.$$  Then,  one can build up  a matrix eigenfunction  ${\bf \Psi} \lf[ N\ri]$
 belonging to \SLd and, 
by developping in the chosen 
basis of the algebra the quantity 
${\bf \Psi}^{-1} \lf[ N\ri] \tau_{0} {\bf \Psi} 
\lf[ N\ri]$, one finds  solutions  for the spin system \eqr{HM}. 
Finally, by using \eqr{metricHM} (or \eqr{Smetric} ) on
\par For the model \eqr{readiff} we can write the  analogue of the dark
soliton, i.e. 
\eq q^{\mp}(x,t) = k e^{\mp \lf[({1
\over 4}v^{2} + 2 k^{2})t + {1 \over 2}v \lf(\xi -\xi_{0}\ri)\ri]}
\tanh \; k \xi , \en which   becomes a time
independent   kink  for $v^{2} = 8k^{2}$.  In contrast with the
  case \eqr{brisol}, now 
   it is impossible to choose the parameters in order to have vanishing
boundary
 conditions at both infinities. \par Finally, looking for homogeneous mass
density solutions such that 
$ q^{+}q^{-} = {1 \over 2} \dot \chi (t)$ holds for any regular function $\chi
\lf( t \ri)$, one finds  
\eq q^{\pm }(x,t) = 
{{c_{\pm} e^{ \mp \chi_{0}}} \over {(t - t_{0})^{{1 \over 2} \pm
          2 c_{+}c_{-}}}}e^{\mp {x^{2} \over 4(t - t_{0})}}
 \virg\en where $c_{\pm}$, $\chi_{0}$ and $t_{0}$ are arbitrary reals.
 This solution
    describes the nonlinear deformation of the Wiener solution for the heat
equation with   $q^{+}q^{-} = {c_{+}c_{-} \over {t - t_{0}}} $.
\par In conclusion we would  like to stress some points. First of all, the same
scheme can be used in embedding into the Jackiw-Teitelboim model other
completely integrable spin models, like the conformal invariant
$\sigma$-model for the classical spin vector ${\bf s}$,
\eq\partial _{+}\partial _{-}{\bf s} - (\partial _{+}{\bf s}, \partial _{-}{\bf
s}){\bf s} = 0\pu\label{confsigma}\en After some manipulations on the
abelian gauge field, also in this case one is led to the  conformal invariant
hyperbolic Sinh-Gordon equation
\eq \partial_{+}\partial_{-}\phi = 2(e^{\phi} - U_{+}U_{-}e^{-\phi}),
\en  where we have introduced  the derivatives in the light-cone coordinates
$\partial_{\pm} = \partial_{2} \pm \partial_{1}$,  $  e^{\phi} =
q^{+}_{+}q^{-}_{-}$ and the arbitrary chiral functions $U_{\pm} =
q^{+}_{\pm}q^{-}_{\pm}$ ( all $ X^{\pm}_{\pm}$ quantities are given by
analogous light-cone linear combinations). The tangent space representation
of the  gauge condition 
\eqr{confsigma} is given by $D^{-}_{-}q^{+}_{+} = 0, D^{+}_{-}q^{-}_{+} = 0 $,
in analogy with  \eqr{constr}. Properly choosing 
$U_{\pm}$, one is lead to the Liouville equation or to conformally broken
completely integrable equations, like the Sinh-Gordon and Cosh-Gordon
equations. The model \eqr{confsigma} is of  its own interest  and a detailed
analysis has been performed, but it is out from the purposes of this work.
Indeed, here we want to outline the appearance of dissipative structures in
gravity theory in the sense expressed by Eq. \eqr{readiff}. This system is
very similar to the "fictitious" or  "mirror-image" systems
 with $negative$ friction, which appear  into  the termo-field  approach to
the  damped harmonic oscillator treated in
\cite{Cel}. 
The energy  is drained from the ``real''  oscillator to its ``image'',
which mimics inaccessible states hidden in a thermostate. In this way the
total energy is conserved
  and the Lagrangian description is allowed. That all this has some implication
on certain thermodynamical properties of the gravity in the presence of
black-holes \cite{Haw,Isr}  is questionable, but suggestive. \par
\par Furthermore, inspired by the gauge equivalence between integrable
models and related
$\sigma$ models, the idea of physical interpretation for gauge degrees
 of freedom can be useful. In fact,  recently in \cite{Kohler} the analogy
between point particles in 2+1 dimensional gravity and linear defects in solid
   continua was reconsidered in the framework of the  CS  gauge theory.
In this analogy the essential difference arises from the role
  of diffeomorphisms which are gauge
 degrees of freedom (unobservable) in gravity,
 while in crystal physics they constitute elastic
  deformations, providing  stress fields in the lattice. 
   \subsection{ Aknowledgments} This work was supported in part by MURST
of Italy
 and by INFN - Sezione di Lecce. One of the authors (O. K. P.) thanks
      the Department of Physics of Lecce University for the warm
   hospitality and A.  Fillipov, A.  Pogrebkov and G. Vitiello for
very useful discussions.
  \vfill
                               \newpage

\end{document}